\documentclass{ws-p9-75x6-50}

\begin{document}

\title{Morphology of the Large-Scale Structure}

\author{Alvaro Dominguez}

\address{Ludwig--Maximilians--Universit\"at, Theresienstr. 37, D-80333
  M\"unchen, Germany \\ E-mail: alvaro@theorie.physik.uni-muenchen.de}

\maketitle

\abstracts{The Minkowski functionals are a mathematical tool to
  quantify morphological features of patterns.  Some applications to
  the matter distribution in galaxy catalogues and N-body simulations
  are reviewed, with an emphasis on the effects of cosmic variance.
  The conclusions are that (i) the observed large-scale morphology is
  sensitive to cosmic variance on scales much larger than the
  nonlinear length ($\approx 8\ h^{-1}$Mpc), and (ii) the large-scale
  morphology predicted by simulations is thus affected by finite-size
  effects, but nonetheless a $\Lambda$CDM model is favored.}

\section{The Minkowski functionals}

The Minkowski functionals $\{M_\mu (B)\}$ are a set of morphological
descriptors of the convex body $B$. There are $d+1$ Minkowski
functionals in the $d$-dimensional Euclidean space (see
Table~\ref{tabmin} for the geometrical interpretation in $d=3$). What
makes these functionals interesting is that they build a {\em
  complete} set of morphological descriptors in the following sense
(Hadwiger's theorem~\cite{KlRo97}): Let $M (B)$ be any functional of
the bodies $B$ in the polyconvex ring (the set of all finite unions of
convex bodies) and let $M(B)$ satisfy properties (a-c) below. Then,
this functional can be written as a linear combination of the
Minkowski functionals. The properties to be satisfied by $M(B)$ are
rather general and apparently little restrictive:
\begin{itemize}
\item[(a)] Motion invariance: $M (B) = M (g B)$ ($g=$ rotation+translation).
\item[(b)] Additivity: $M (B_1\cup B_2) = M (B_1) + M (B_2)- M (B_1\cap B_2)$.
\item[(c)] Convex continuity: $M (B_i) \longrightarrow M (B)$ as $B_i
  \longrightarrow B$ ($B, B_i$ {\em convex}).
\end{itemize}

\begin{table}[b]
  \caption{Definition of the Minkowski functionals $M_\mu$ in 3-dimensional 
    Euclidean space \label{tabmin}}
  \begin{center}
    \begin{tabular}{cl|c|c}
      & geometric quantity      & $\mu$ & $M_\mu$      \\ 
      \hline 
      & & & \\
      $V$ & volume                  & 0     & $V$          \\
      $A$ & surface                 & 1     & $A/8$        \\
      $H$ & integrated surface mean curvature & 2     & $H/2\pi^2$   \\
      $\chi$ & Euler characteristic & 3     & $3\chi/4\pi$ \\
    \end{tabular}
  \end{center}
\end{table}

\section{Cosmological applications of the Minkowski functionals}

\begin{figure}[t]
  \begin{center}
    \begin{minipage}[l]{9pc}
      \epsfxsize=9pc
      \epsfbox{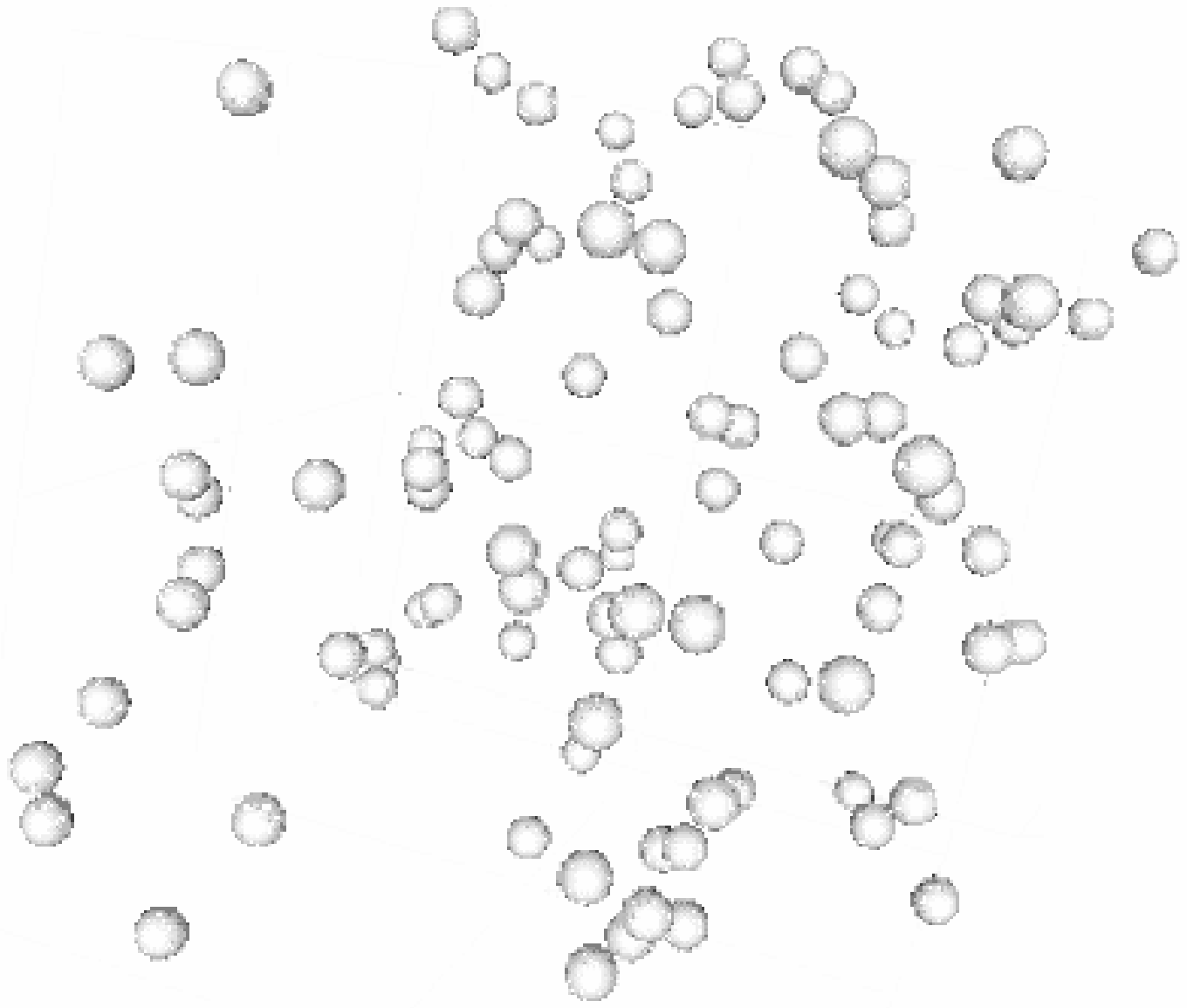}
    \end{minipage}
    \begin{minipage}[c]{9pc}
      \epsfxsize=9pc
      \epsfbox{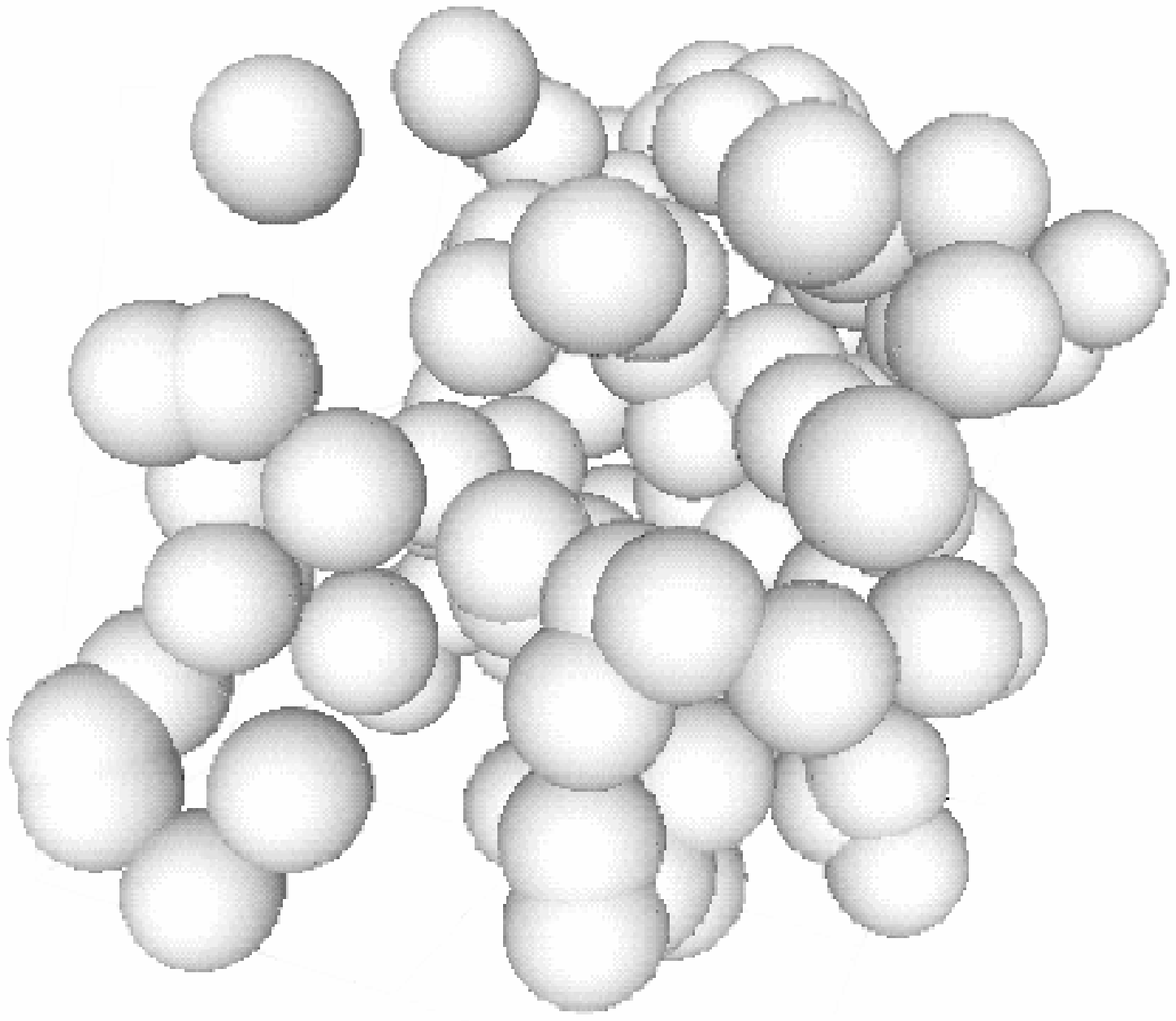}
    \end{minipage}
    \begin{minipage}[r]{9pc}
      \epsfxsize=9pc
      \epsfbox{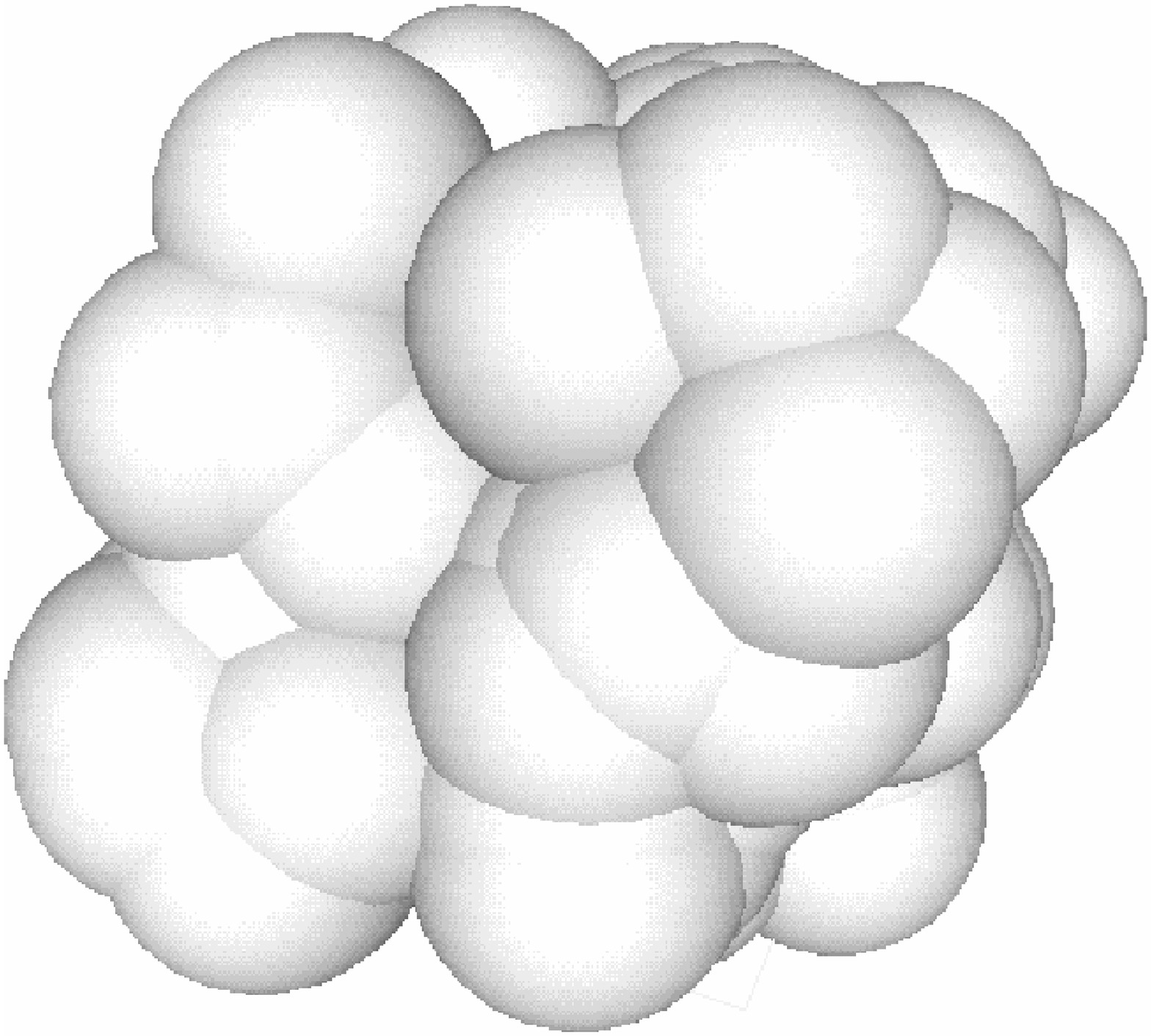}
    \end{minipage}
  \end{center}
  \caption{
    The Boolean grain model with varying ball radius.
    \label{figboolean}}
\end{figure}

The Minkowski functionals have been only recently applied to quantify
the morphological properties of the galaxy
distribution~\cite{MBW94,Kers99}. Galaxy catalogues provide a
distribution of points; a simple method to associate a body to the
point distribution is the {\it Boolean grain model}: every point is
taken to be the center of a ball of a given radius $r$, and one gets a
body ${\cal B}_r$ formed by the union of all the balls
(Fig.~\ref{figboolean}).  ${\cal B}_r$ clearly belongs to the
polyconvex ring and so Hadwiger's theorem applies.

For models of stochastic point processes, one can consider the
ensemble average of the Minkowski functionals as a function of the
ball radius, $\langle M_\mu \rangle (r)$. These averages depend on all
the correlation functions of the stochastic
process~\cite{MBW94,SGKK99}; hence, the Minkowski functionals may be
useful in cosmological investigations as a compact way for collecting
information beyond the two-point level. As a matter of fact, some of
the Minkowski functionals had already been employed in the
cosmological literature: $\langle M_0 \rangle$ is essentially the void
probability function, while $\langle M_3 \rangle$ is equivalent to the
genus analysis. And the four Minkowski functionals share with these by
now standard cosmological tools the robustness against errors.

A Poissonian point process (no correlations at all) provides a simple
reference model for comparison, since the average Minkowski
functionals can be computed analytically~\cite{MeWa91}. The shaded
area in Fig.~\ref{figabell} represents the volume density of the
average Minkowski functionals ($m_\mu$) for this Poissonian case
together with the region of 1-$\sigma$ variations. For small radii,
the balls barely overlap and $m_\mu$ corresponds to the Minkowski
functional of an isolated ball. For large radii, the balls intersect
to yield a single big body: $m_0$ saturates and the other $m_\mu$
become very small. The maximum in $m_1$ is due to the decrease in the
exposed area of the balls as they start overlapping. For the same
reason, $m_2$ has first a maximum but then becomes negative, due to
the dominant negative contributions of the intersection arcs between
pairs of balls.  Finally, $m_3$ exhibits first a negative minimum due
to the tunnels in the structure (negative contribution to $\chi$), but
then there is a positive maximum because of the cavities formed by the
closing tunnels (positive contribution to $\chi$).

\subsection{Galaxy catalogues}

\begin{figure}[t]
  \begin{center}
    \epsfxsize=20pc 
    \epsfbox{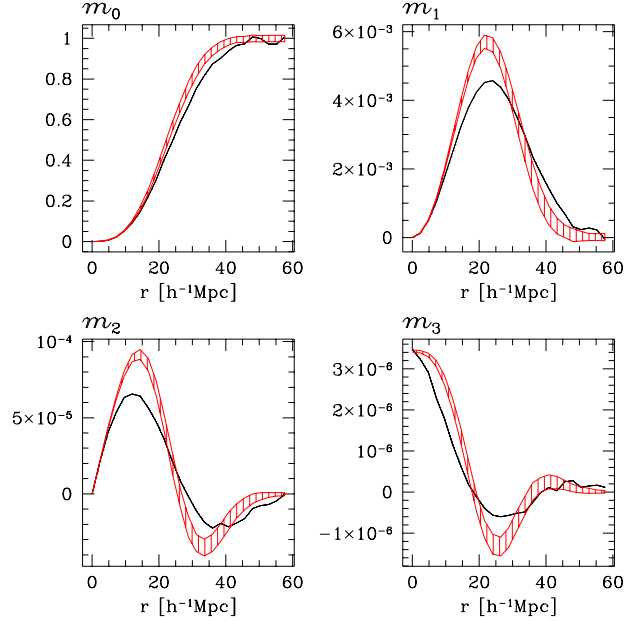}
    \caption{
      Minkowski functionals for the Abell/ACO cluster catalogue (solid
      line) and for a Poissonian point process with the same number
      density (shaded region, encompassing 1-$\sigma$ deviations).
      \label{figabell}}
  \end{center}
\end{figure}

Fig.~\ref{figabell} shows the Minkowski functionals for a
volume-limited sample ($\approx 240\ h^{-1}$Mpc deep) extracted from
the Abell/ACO cluster catalogue~\cite{Ketal97}, containing $\approx
400$ clusters. Also shown are the functionals of a Poissonian point
process with the same number density. A significant deviation from the
Poissonian case is observed: from the interpretation of the main
features of the Poissonian curves, one concludes that this deviation
describes an enhanced clustering on scales $\sim 15$-$50\ h^{-1}$Mpc.

\begin{figure}[t]
  \begin{center}
    \begin{minipage}[l]{9pc}
      \epsfxsize=9pc
      \epsfbox{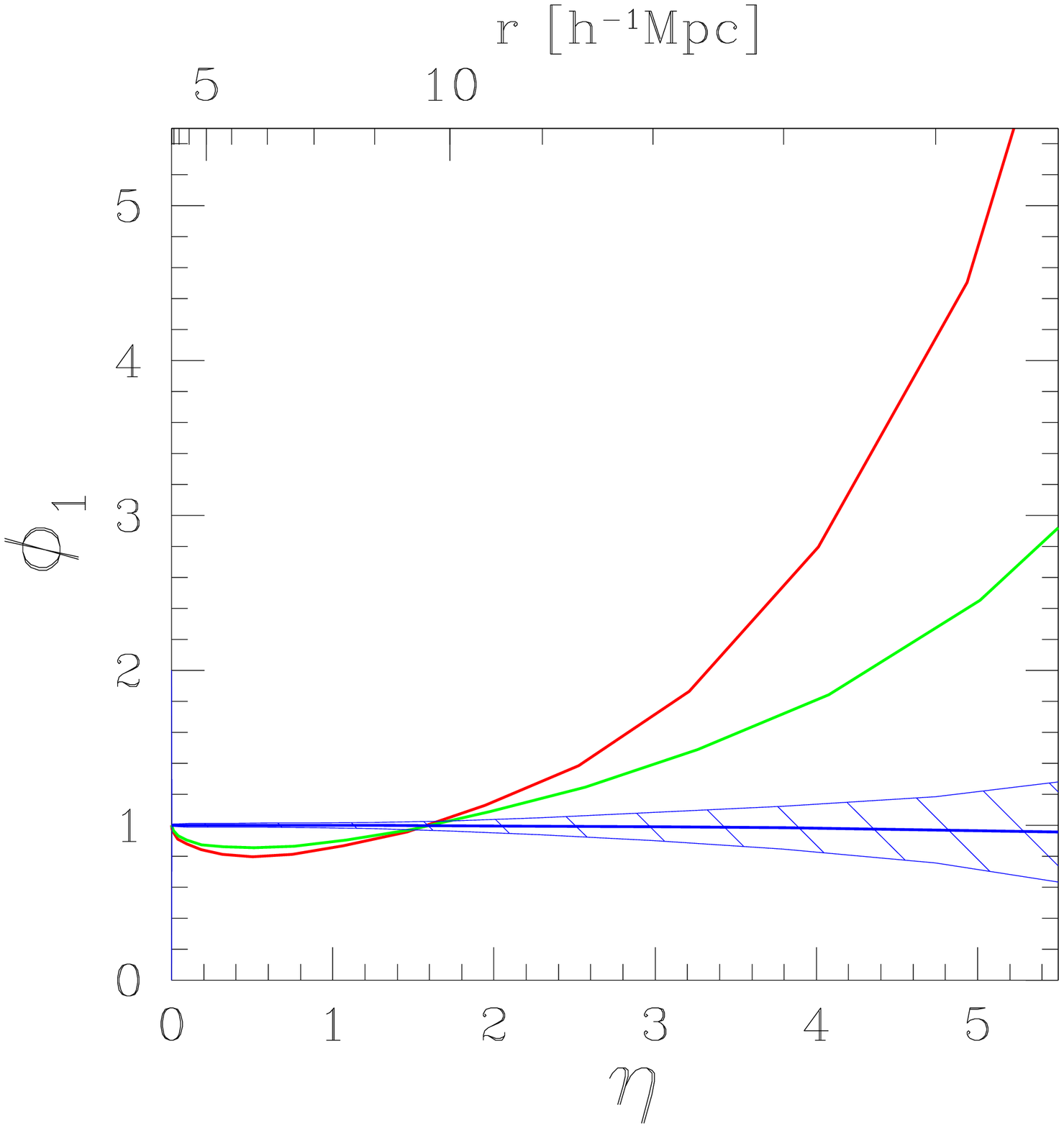}
    \end{minipage}
    \begin{minipage}[c]{9pc}
      \epsfxsize=9pc
      \epsfbox{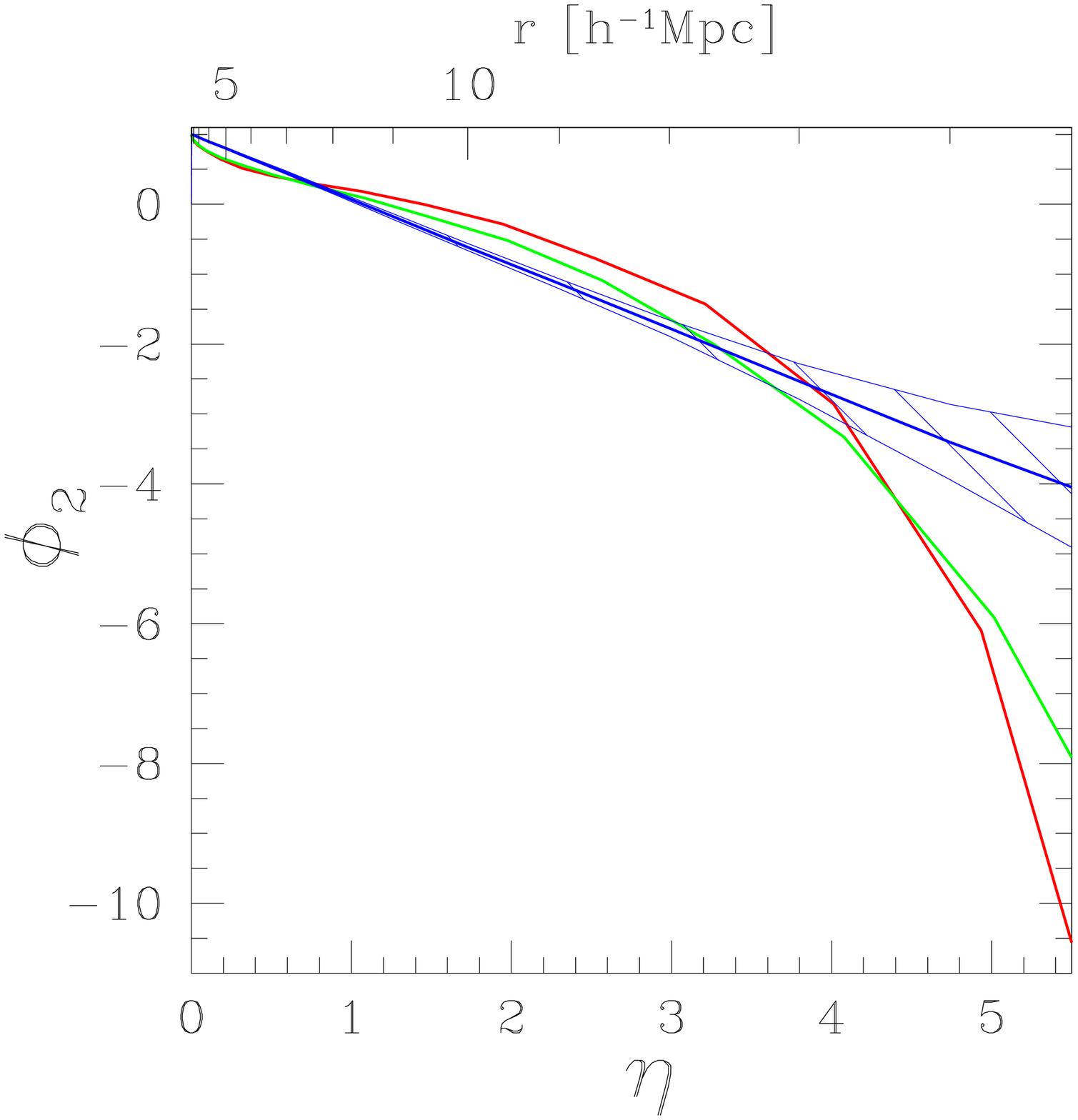}
    \end{minipage}
    \begin{minipage}[r]{9pc}
      \epsfxsize=9pc
      \epsfbox{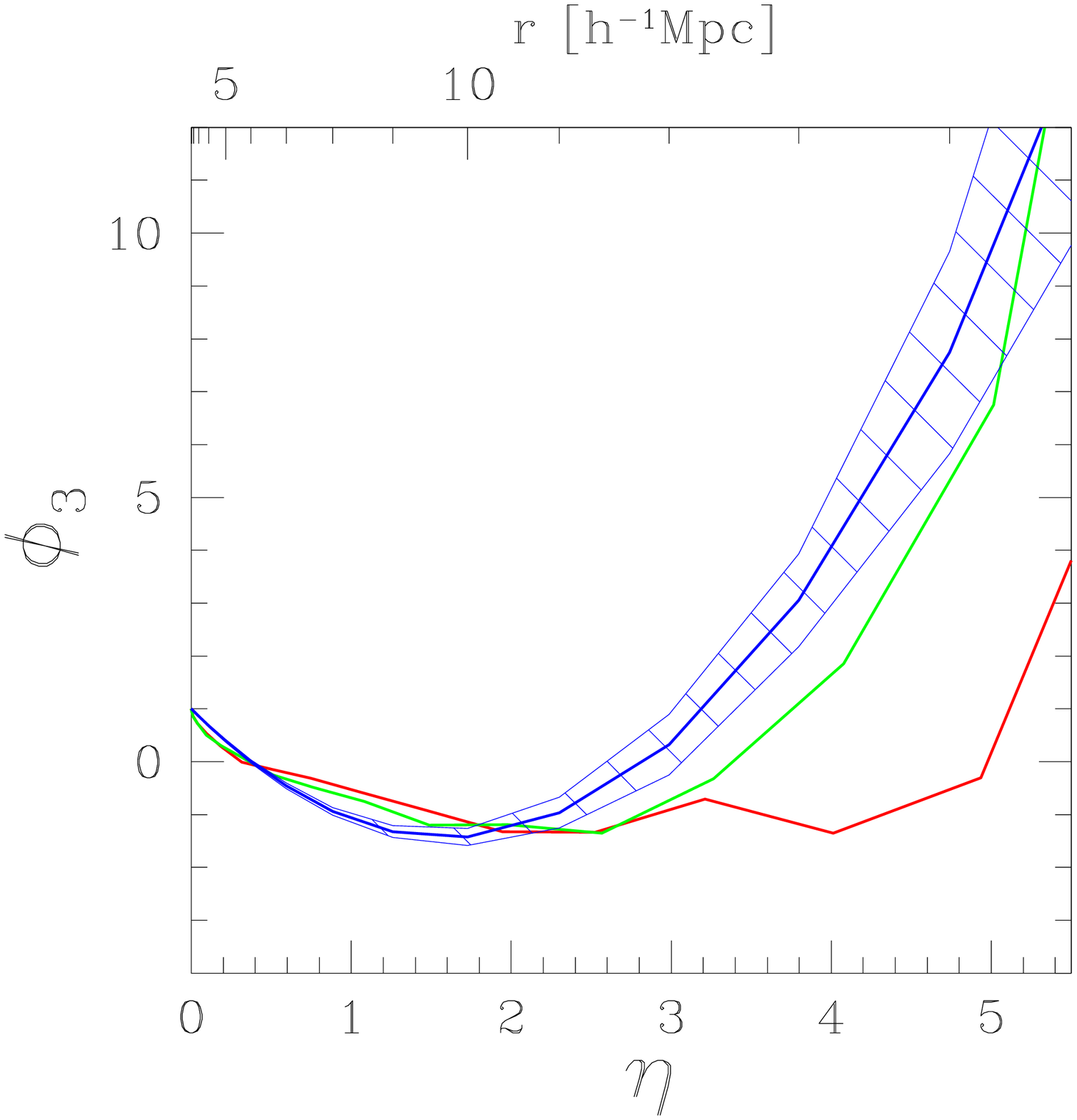}
    \end{minipage}
  \end{center}
  \caption{
    Minkowski functionals for the northern and southern (darker line)
    parts of the PSCz galaxy catalogue and for a Poissonian point
    process with the same number density (shaded region: 1-$\sigma$
    deviations). Deviations from the Poissonian case are larger for
    the southern part. \label{figpscz}}
\end{figure}

Fig.~\ref{figpscz} shows the last three Minkowski functionals for a
volume-limited sample ($\approx 100\ h^{-1}$Mpc deep) extracted from
the PSCz galaxy catalogue~\cite{Ketal01}, containing $\approx 1300$
galaxies. The Minkowski functionals were normalized by $\langle M_1
\rangle_{\rm Poisson}$ in order to enhance the visibility. The
Minkowski functionals of samples from the IRAS catalogue~\cite{KSBW98}
(covering the same volume but less complete than the samples from the
PSCz catalogue) exhibit the same behavior. As well as enhanced
clustering, a significant deviation in the morphology between the
northern and southern parts is detected. By comparing the Minkowski
functionals of subsamples within the northern part and within the
southern part, respectively, it could be checked that this deviation
does not arise from a particular North-South anisotropy, but it is
rather a signature of cosmic variance on the large scales of the
sample. In fact, this variance persists when the sample depth is
stretched till $200\ h^{-1}$Mpc.

The same kind of analysis was repeated for the CfA2
catalogue~\cite{ScDi00}. A significant deviation in the morphological
features between North and South is again observed. But this
conclusion is not so robust, because the sample depth, $\approx 70\ 
h^{-1}$Mpc, is too small: the maximum reliable ball radius is $5\ 
h^{-1}$Mpc and the results are likely dominated by local effects, as
$70\ h^{-1}$Mpc-deep subsamples of the IRAS catalogue point
out~\cite{KSBW98,ScDi00}.

In conclusion, the morphological features of the galaxy distribution,
as measured by the Minkowski functionals, show the effects of cosmic
variance on scales larger than the ``nonlinear scale'' $r_0 \approx 8\ 
h^{-1}$Mpc (defined through $\sigma^2 (r_0) = \langle \delta^2 \rangle
=1$). This must be a consequence of the sensitivity of the Minkowski
functionals to correlations beyond the two-point level. In fact, they
can be used to quantify the degree of planarity and filamentarity of a
pattern~\cite{ScDi00,Setal99}. So, e.g., it could be checked that
galaxies in the CfA2 catalogue tend to be distributed in planar
structures: this is likely a signature of the ``Great Wall'' and a
major reason for the observed cosmic variance in this catalogue.

\subsection{N-body simulations}

\begin{figure}[t]
  \begin{center}
    \epsfxsize=20pc
    \epsfbox{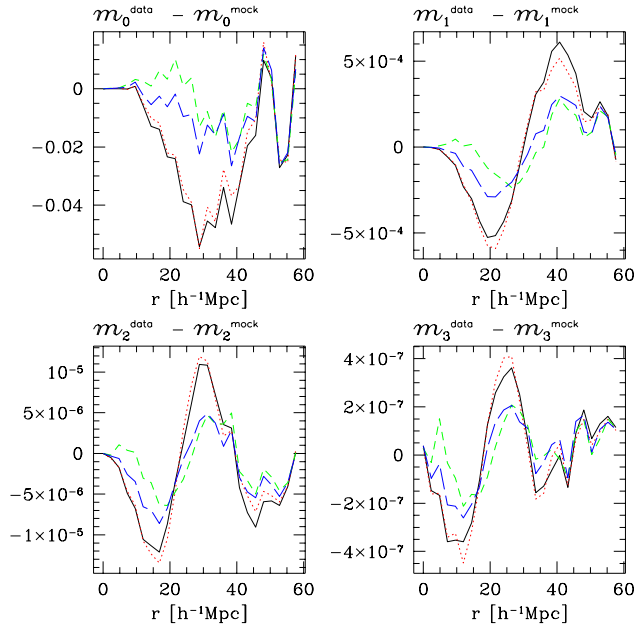}
  \end{center}
  \caption{
    Difference between the Minkowski functionals of the Abell/ACO
    catalogue and those of mock catalogues for four cosmological
    models: CDM (solid), tCDM (dotted), $\Lambda$CDM (long dashed),
    BSI (short dashed). \label{figsimul}}
\end{figure}

Fig.~\ref{figsimul} shows the Minkowski functionals extracted from
cluster mock catalogues and compared with those from the Abell/ACO
cluster catalogue~\cite{Ketal97}. The mock catalogues correspond to
four different cosmological models (side length of the simulation box
$= 500\ h^{-1}$Mpc): standard CDM, tilted CDM (tCDM), $\Lambda$CDM,
and a model with broken scale invariance (BSI) in the primordial
spectrum.  The deviations of the CDM and tCDM predictions from
observations are larger than the estimated 1-$\sigma$ dispersion (not
shown) in the direction towards less clustering. The $\Lambda$CDM and
BSI models fit the data much better, but they still exhibit a tendency
towards less structure.

The Minkowski functionals of the CfA2 catalogue have also been
compared with mock catalogues for two cosmological models:
$\Lambda$CDM (box side length $=141\ h^{-1}$Mpc) and $\tau$CDM (box
side length $=85\ h^{-1}$Mpc)~\cite{ScDi00}. The models predict again
less clustering than observed, and, within this condition, the
$\Lambda$CDM model is slightly favored.

A first conclusion is therefore the inability of the N-body
simulations to reproduce the observed large-scale morphology of the
galaxy distribution. Simulations systematically predict less structure
than observed. A possible explanation for this result emerges when it
is combined with the effects of cosmic variance mentioned in the
previous subsection: the simulation volumes are too small to address
the question of the large-scale morphology, so that the Minkowski
functionals derived from the simulations suffer from finite-volume
corrections. Within this cautionary remark, a second conclusion is
that a cosmological $\Lambda$CDM model seems to provide the best fit
to the observed large-scale morphology.

\section*{Acknowledgments}

The author thanks C. Beisbart, T. Buchert, M. Kerscher, J. Schmalzing
and H. Wagner for their help and for permission to reproduce the
figures. He also thanks the organizers for the invitation to
participate in the workshop.

\end{document}